\overfullrule=0pt
%
%
%
%
%
%
\def\unredoffs{} 

%
%
%
%
\newbox\leftpage \newdimen\fullhsize \newdimen\hstitle \newdimen\hsbody
\tolerance=1000\hfuzz=2pt
\catcode`\@=11 
\def\bigans{b }
%
\magnification=1200\unredoffs\baselineskip=16pt plus 2pt minus 1pt
\hsbody=\hsize \hstitle=\hsize 
%
%
%
\newcount\yearltd\yearltd=\year\advance\yearltd by -1900

\def\Title#1#2{\nopagenumbers\abstractfont\hsize=\hstitle\rightline{#1}%
\vskip 1in\centerline{\titlefont #2}\abstractfont\vskip .5in\pageno=0}
\def\Date#1{\vfill\leftline{#1}\tenpoint\supereject\global\hsize=\hsbody%
\footline={\hss\tenrm\folio\hss}}
%

\def\draftmode{\message{ DRAFTMODE }\def\draftdate{{\rm preliminary draft:
\number\month/\number\day/\number\yearltd\ \ \hourmin}}%
\headline={\hfil\draftdate}\writelabels\baselineskip=20pt plus 2pt minus 2pt
 {\count255=\time\divide\count255 by 60 \xdef\hourmin{\number\count255}
  \multiply\count255 by-60\advance\count255 by\time
  \xdef\hourmin{\hourmin:\ifnum\count255<10 0\fi\the\count255}}}
\def\nolabels{\def\wrlabeL##1{}\def\eqlabeL##1{}\def\reflabeL##1{}}
\def\writelabels{\def\wrlabeL##1{\leavevmode\vadjust{\rlap{\smash%
{\line{{\escapechar=` \hfill\rlap{\sevenrm\hskip.03in\string##1}}}}}}}%
\def\eqlabeL##1{{\escapechar-1\rlap{\sevenrm\hskip.05in\string##1}}}%
\def\reflabeL##1{\noexpand\llap{\noexpand\sevenrm\string\string\string##1}}}
\nolabels
%
\global\newcount\secno \global\secno=0
\global\newcount\meqno \global\meqno=1
\def\newsec#1{\global\advance\secno by1\message{(\the\secno. #1)}
\global\subsecno=0\eqnres@t\noindent{\bf\the\secno. #1}
\writetoca{{\secsym} {#1}}\par\nobreak\medskip\nobreak}
\def\eqnres@t{\xdef\secsym{\the\secno.}\global\meqno=1\bigbreak\bigskip}
\def\sequentialequations{\def\eqnres@t{\bigbreak}}\xdef\secsym{}
\global\newcount\subsecno \global\subsecno=0
\def\subsec#1{\global\advance\subsecno by1\message{(\secsym\the\subsecno. #1)}
\ifnum\lastpenalty>9000\else\bigbreak\fi
\noindent{\it\secsym\the\subsecno. #1}\writetoca{\string\quad 
{\secsym\the\subsecno.} {#1}}\par\nobreak\medskip\nobreak}
\def\appendix#1#2{\global\meqno=1\global\subsecno=0\xdef\secsym{\hbox{#1.}}
\bigbreak\bigskip\noindent{\bf Appendix #1. #2}\message{(#1. #2)}
\writetoca{Appendix {#1.} {#2}}\par\nobreak\medskip\nobreak}
%
%
\def\eqnn#1{\xdef #1{(\secsym\the\meqno)}\writedef{#1\leftbracket#1}%
\global\advance\meqno by1\wrlabeL#1}
\def\eqna#1{\xdef #1##1{\hbox{$(\secsym\the\meqno##1)$}}
\writedef{#1\numbersign1\leftbracket#1{\numbersign1}}%
\global\advance\meqno by1\wrlabeL{#1$\{\}$}}
\def\eqn#1#2{\xdef #1{(\secsym\the\meqno)}\writedef{#1\leftbracket#1}%
\global\advance\meqno by1$$#2\eqno#1\eqlabeL#1$$}
%
\newskip\footskip\footskip14pt plus 1pt minus 1pt 
\def\footnotefont{\ninepoint}\def\f@t#1{\footnotefont #1\@foot}
\def\f@@t{\baselineskip\footskip\bgroup\footnotefont\aftergroup\@foot\let\next}
\setbox\strutbox=\hbox{\vrule height9.5pt depth4.5pt width0pt}
\global\newcount\ftno \global\ftno=0
\def\foot{\global\advance\ftno by1\footnote{$^{\the\ftno}$}}
%
\newwrite\ftfile   
\def\footend{\def\foot{\global\advance\ftno by1\chardef\wfile=\ftfile
$^{\the\ftno}$\ifnum\ftno=1\immediate\openout\ftfile=foots.tmp\fi%
\immediate\write\ftfile{\noexpand\smallskip%
\noexpand\item{f\the\ftno:\ }\pctsign}\findarg}%
\def\footatend{\vfill\eject\immediate\closeout\ftfile{\parindent=20pt
\centerline{\bf Footnotes}\nobreak\bigskip\input foots.tmp }}}
\def\footatend{}
%
%
\global\newcount\refno \global\refno=1
\newwrite\rfile
\def\ref{[\the\refno]\nref}
\def\nref#1{\xdef#1{[\the\refno]}\writedef{#1\leftbracket#1}%
\ifnum\refno=1\immediate\openout\rfile=refs.tmp\fi
\global\advance\refno by1\chardef\wfile=\rfile\immediate
\write\rfile{\noexpand\item{#1\ }\reflabeL{#1\hskip.31in}\pctsign}\findarg}
\def\findarg#1#{\begingroup\obeylines\newlinechar=`\^^M\pass@rg}
{\obeylines\gdef\pass@rg#1{\writ@line\relax #1^^M\hbox{}^^M}%
\gdef\writ@line#1^^M{\expandafter\toks0\expandafter{\striprel@x #1}%
\edef\next{\the\toks0}\ifx\next\em@rk\let\next=\endgroup\else\ifx\next\empty%
\else\immediate\write\wfile{\the\toks0}\fi\let\next=\writ@line\fi\next\relax}}
\def\striprel@x#1{} \def\em@rk{\hbox{}} 
\def\lref{\begingroup\obeylines\lr@f}
\def\lr@f#1#2{\gdef#1{\ref#1{#2}}\endgroup\unskip}
\def\semi{;\hfil\break}
\def\addref#1{\immediate\write\rfile{\noexpand\item{}#1}} 
\def\startrefs#1{\immediate\openout\rfile=refs.tmp\refno=#1}
\def\xref{\expandafter\xr@f}\def\xr@f[#1]{#1}
\def\refs#1{\count255=1[\r@fs #1{\hbox{}}]}
\def\r@fs#1{\ifx\und@fined#1\message{reflabel \string#1 is undefined.}%
\nref#1{need to supply reference \string#1.}\fi%
\vphantom{\hphantom{#1}}\edef\next{#1}\ifx\next\em@rk\def\next{}%
\else\ifx\next#1\ifodd\count255\relax\xref#1\count255=0\fi%
\else#1\count255=1\fi\let\next=\r@fs\fi\next}
%

%
\newwrite\ffile\global\newcount\figno \global\figno=1
\def\fig{fig.~\the\figno\nfig}
\def\nfig#1{\xdef#1{fig.~\the\figno}%
\writedef{#1\leftbracket fig.\noexpand~\the\figno}%
\ifnum\figno=1\immediate\openout\ffile=figs.tmp\fi\chardef\wfile=\ffile%
\immediate\write\ffile{\noexpand\medskip\noexpand\item{Fig.\ \the\figno. }
\reflabeL{#1\hskip.55in}\pctsign}\global\advance\figno by1\findarg}
\def\xfig{\expandafter\xf@g}\def\xf@g fig.\penalty\@M\ {}
\def\figs#1{figs.~\f@gs #1{\hbox{}}}
\def\f@gs#1{\edef\next{#1}\ifx\next\em@rk\def\next{}\else
\ifx\next#1\xfig #1\else#1\fi\let\next=\f@gs\fi\next}
\newwrite\lfile
{\escapechar-1\xdef\pctsign{\string\%}\xdef\leftbracket{\string\{}
\xdef\rightbracket{\string\}}\xdef\numbersign{\string\#}}

\def\writestop{\def\writestoppt{\immediate\write\lfile{\string\pageno%
\the\pageno\string\startrefs\leftbracket\the\refno\rightbracket%
\string\def\string\secsym\leftbracket\secsym\rightbracket%
\string\secno\the\secno\string\meqno\the\meqno}\immediate\closeout\lfile}}
\def\writestoppt{}\def\writedef#1{}
\def\seclab#1{\xdef #1{\the\secno}\writedef{#1\leftbracket#1}\wrlabeL{#1=#1}}
\def\subseclab#1{\xdef #1{\secsym\the\subsecno}%
\writedef{#1\leftbracket#1}\wrlabeL{#1=#1}}
\newwrite\tfile \def\writetoca#1{}
\def\leaderfill{\leaders\hbox to 1em{\hss.\hss}\hfill}
\def\writetoc{\immediate\openout\tfile=toc.tmp 
   \def\writetoca##1{{\edef\next{\write\tfile{\noindent ##1 
   \string\leaderfill {\noexpand\number\pageno} \par}}\next}}}

%
\catcode`\@=12 
%
\edef\tfontsize{\ifx\answ\bigans scaled\magstep3\else scaled\magstep4\fi}
\font\titlerm=cmr10 \tfontsize \font\titlerms=cmr7 \tfontsize
\font\titlermss=cmr5 \tfontsize \font\titlei=cmmi10 \tfontsize
\font\titleis=cmmi7 \tfontsize \font\titleiss=cmmi5 \tfontsize
\font\titlesy=cmsy10 \tfontsize \font\titlesys=cmsy7 \tfontsize
\font\titlesyss=cmsy5 \tfontsize \font\titleit=cmti10 \tfontsize
\skewchar\titlei='177 \skewchar\titleis='177 \skewchar\titleiss='177
\skewchar\titlesy='60 \skewchar\titlesys='60 \skewchar\titlesyss='60
\def\titlefont{\def\rm{\fam0\titlerm}
\textfont0=\titlerm \scriptfont0=\titlerms \scriptscriptfont0=\titlermss
\textfont1=\titlei \scriptfont1=\titleis \scriptscriptfont1=\titleiss
\textfont2=\titlesy \scriptfont2=\titlesys \scriptscriptfont2=\titlesyss
\textfont\itfam=\titleit \def\it{\fam\itfam\titleit}\rm}
 \ifx\answ\bigans\else scaled\magstep1\fi
\ifx\answ\bigans\def\abstractfont{\tenpoint}\else
\font\abssl=cmsl10 scaled \magstep1
\font\absrm=cmr10 scaled\magstep1 \font\absrms=cmr7 scaled\magstep1
\font\absrmss=cmr5 scaled\magstep1 \font\absi=cmmi10 scaled\magstep1
\font\absis=cmmi7 scaled\magstep1 \font\absiss=cmmi5 scaled\magstep1
\font\abssy=cmsy10 scaled\magstep1 \font\abssys=cmsy7 scaled\magstep1
\font\abssyss=cmsy5 scaled\magstep1 \font\absbf=cmbx10 scaled\magstep1
\skewchar\absi='177 \skewchar\absis='177 \skewchar\absiss='177
\skewchar\abssy='60 \skewchar\abssys='60 \skewchar\abssyss='60
\def\abstractfont{\def\rm{\fam0\absrm}
\textfont0=\absrm \scriptfont0=\absrms \scriptscriptfont0=\absrmss
\textfont1=\absi \scriptfont1=\absis \scriptscriptfont1=\absiss
\textfont2=\abssy \scriptfont2=\abssys \scriptscriptfont2=\abssyss
\textfont\itfam=\bigit \def\it{\fam\itfam\bigit}\def\footnotefont{\tenpoint}%
\textfont\slfam=\abssl \def\sl{\fam\slfam\abssl}%
\textfont\bffam=\absbf \def\bf{\fam\bffam\absbf}\rm}\fi
\def\tenpoint{\def\rm{\fam0\tenrm}
\textfont0=\tenrm \scriptfont0=\sevenrm \scriptscriptfont0=\fiverm
\textfont1=\teni  \scriptfont1=\seveni  \scriptscriptfont1=\fivei
\textfont2=\tensy \scriptfont2=\sevensy \scriptscriptfont2=\fivesy
\textfont\itfam=\tenit \def\it{\fam\itfam\tenit}\def\footnotefont{\ninepoint}%
\textfont\bffam=\tenbf \def\bf{\fam\bffam\tenbf}\def\sl{\fam\slfam\tensl}\rm}
\font\ninerm=cmr9 \font\sixrm=cmr6 \font\ninei=cmmi9 \font\sixi=cmmi6 
\font\ninesy=cmsy9 \font\sixsy=cmsy6 \font\ninebf=cmbx9 
\font\nineit=cmti9 \font\ninesl=cmsl9 \skewchar\ninei='177
\skewchar\sixi='177 \skewchar\ninesy='60 \skewchar\sixsy='60 
\def\ninepoint{\def\rm{\fam0\ninerm}
\textfont0=\ninerm \scriptfont0=\sixrm \scriptscriptfont0=\fiverm
\textfont1=\ninei \scriptfont1=\sixi \scriptscriptfont1=\fivei
\textfont2=\ninesy \scriptfont2=\sixsy \scriptscriptfont2=\fivesy
\textfont\itfam=\ninei \def\it{\fam\itfam\nineit}\def\sl{\fam\slfam\ninesl}%
\textfont\bffam=\ninebf \def\bf{\fam\bffam\ninebf}\rm} 
%
%

\hyphenation{anom-aly anom-alies coun-ter-term coun-ter-terms}
\def\inv{^{\raise.15ex\hbox{${\scriptscriptstyle -}$}\kern-.05em 1}}

\def\Dsl{\,\raise.15ex\hbox{/}\mkern-13.5mu D} 
\def\dsl{\raise.15ex\hbox{/}\kern-.57em\partial}

\font\bigit=cmti10 scaled \magstep1
\def\lspace{\ifx\answ\bigans{}\else\qquad\fi}
\def\lbspace{\ifx\answ\bigans{}\else\hskip-.2in\fi} 
\def\boxeqn#1{\vcenter{\vbox{\hrule\hbox{\vrule\kern3pt\vbox{\kern3pt
	\hbox{${\displaystyle #1}$}\kern3pt}\kern3pt\vrule}\hrule}}}
\def\mbox#1#2{\vcenter{\hrule \hbox{\vrule height#2in
		\kern#1in \vrule} \hrule}}  
%

\def\e#1{{\rm e}^{^{\textstyle#1}}}

\def\darr#1{\raise1.5ex\hbox{$\leftrightarrow$}\mkern-16.5mu #1}

\def\half{{\textstyle{1\over2}}} 
\def\roughly#1{\raise.3ex\hbox{$#1$\kern-.75em\lower1ex\hbox{$\sim$}}}

\lref\berksimp{
N.~Berkovits,
``Simplifying and Extending the AdS(5) x S**5 Pure Spinor Formalism,''
JHEP {\bf 09}, 051 (2009)
[arXiv:0812.5074 [hep-th]].}

\lref\berkpert{
N.~Berkovits,
``Perturbative Super-Yang-Mills from the Topological AdS(5) x S**5 Sigma Model,''
JHEP {\bf 09}, 088 (2008)
[arXiv:0806.1960 [hep-th]].}

\lref\berkvafa{
N.~Berkovits and C.~Vafa,
``Towards a Worldsheet Derivation of the Maldacena Conjecture,''
JHEP {\bf 03}, 031 (2008)
[arXiv:0711.1799 [hep-th]].}

\lref\berklimit{
N.~Berkovits,
``A New Limit of the AdS(5) x S**5 Sigma Model,''
JHEP {\bf 08}, 011 (2007)
[arXiv:hep-th/0703282 [hep-th]].}

\lref\berktwistor{
N.~Berkovits,
``An Alternative string theory in twistor space for N=4 superYang-Mills,''
Phys. Rev. Lett. {\bf 93}, 011601 (2004)
[arXiv:hep-th/0402045 [hep-th]].}

\lref\berkads{
N.~Berkovits and O.~Chandia,
``Superstring vertex operators in an AdS(5) x S**5 background,''
Nucl. Phys. B {\bf 596}, 185-196 (2001)
[arXiv:hep-th/0009168 [hep-th]].}

\lref\berkderive{
N.~Berkovits,
``Sketching a Proof of the Maldacena Conjecture at Small Radius,''
JHEP {\bf 06} (2019), 111
[arXiv:1903.08264 [hep-th]].
}

\lref\gopa{
M.~R.~Gaberdiel and R.~Gopakumar,
``String Dual to Free N=4 Supersymmetric Yang-Mills Theory,''
Phys. Rev. Lett. {\bf 127}, no.13, 131601 (2021)
[arXiv:2104.08263 [hep-th]]\semi
M.~R.~Gaberdiel and R.~Gopakumar,
``The worldsheet dual of free super Yang-Mills in 4D,''
JHEP {\bf 11}, 129 (2021)
[arXiv:2105.10496 [hep-th]].}

\lref\gopatwo{
R.~Gopakumar and E.~A.~Mazenc,
``Deriving the Simplest Gauge-String Duality -- I: Open-Closed-Open Triality,''
[arXiv:2212.05999 [hep-th]].
}

\lref\gopavafa{
R.~Gopakumar and C.~Vafa,
``On the gauge theory / geometry correspondence,''
Adv. Theor. Math. Phys. {\bf 3}, 1415-1443 (1999)
[arXiv:hep-th/9811131 [hep-th]].}

\lref\vafa{
H.~Ooguri and C.~Vafa,
``World sheet derivation of a large N duality,''
Nucl. Phys. B {\bf 641}, 3-34 (2002)
[arXiv:hep-th/0205297 [hep-th]].}

\lref\gaiotto{
D.~Gaiotto and L.~Rastelli,
``A Paradigm of open / closed duality: Liouville D-branes and the Kontsevich model,''
JHEP {\bf 07}, 053 (2005)
[arXiv:hep-th/0312196 [hep-th]].}

\lref\grone{
D.~Gaiotto, N.~Itzhaki and L.~Rastelli,
``Closed strings as imaginary D-branes,''
Nucl. Phys. B {\bf 688}, 70-100 (2004)
[arXiv:hep-th/0304192 [hep-th]].}

\lref\wittenb{
E.~Witten,
``Perturbative gauge theory as a string theory in twistor space,''
Commun. Math. Phys. {\bf 252}, 189-258 (2004)
[arXiv:hep-th/0312171 [hep-th]].}

\lref\vafaa{
A.~Neitzke and C.~Vafa,
``N=2 strings and the twistorial Calabi-Yau,''
[arXiv:hep-th/0402128 [hep-th]].}

\lref\howe{
P.~S.~Howe and P.~C.~West,
``Nonperturbative Green's functions in theories with extended superconformal symmetry,''
Int. J. Mod. Phys. A {\bf 14} (1999), 2659-2674
[arXiv:hep-th/9509140 [hep-th]].}

\lref\wittop{
E.~Witten,
``Topological Sigma Models,''
Commun. Math. Phys. {\bf 118}, 411 (1988).}

\lref\phases{
E.~Witten,
``Phases of N=2 theories in two-dimensions,''
Nucl. Phys. B {\bf 403}, 159-222 (1993)
[arXiv:hep-th/9301042 [hep-th]].}

\lref\aisaka{
Y.~Aisaka and Y.~Kazama,
``Operator mapping between RNS and extended pure spinor formalisms for superstring,''
JHEP {\bf 08} (2003), 047
[arXiv:hep-th/0305221 [hep-th]].}

\def\a{{\alpha}}

\def\ad{{\dot a}}

\def\kb{{\bar \kappa}}

\def\l{{\lambda}}
\def\lb{{\overline\lambda}}

\def\lb{{\overline\lambda}}

\def\d{{\delta}}
\def\e{{\epsilon}}
\def\s{{\sigma}}
\def\k{{\kappa}}
\def\kb{{\overline\kappa}}

\def\Ab{{\overline A}}

\def\zb{{\overline z}}

\def\Ib{{\overline I}}

\def\L{{\Lambda}}
\def\Lb{{\overline\Lambda}}

\def\half{{1\over 2}}
\def\p{{\partial}}

\def\pb{{\overline\partial}}
\def\t{{\theta}}
\def\tb{{\overline\theta}}

\def\ad{{\dot \a}}

\def\T{{\Theta}}

\def\S{{\Sigma}}

\def\jb{{\overline j}}

\def\Tb{{\overline\Theta}}
\def\Phib{{\overline \Phi}}

\def\Lb{{\overline\Lambda}}
\def\tb{{\overline\theta}}

\def\Zb{{\overline Z}}

\def\lb{{\overline{\lambda}}}

\Title{\vbox{\baselineskip12pt
\hbox{}}}
{{\vbox{\centerline{Topological A-Model for $AdS_5\times S^5$ Superstring  }
\smallskip
\centerline{ and the Maldacena Conjecture}}} }
\bigskip\centerline{Nathan Berkovits\foot{e-mail: nathan.berkovits@unesp.br}}
\bigskip
\centerline{\it ICTP South American Institute for Fundamental Research}
\centerline{\it Instituto de F\'\i sica Te\'orica, UNESP - Univ. 
Estadual Paulista }
\centerline{\it Rua Dr. Bento T. Ferraz 271, 01140-070, S\~ao Paulo, SP, Brasil}
\bigskip

\vskip .1in

A topological A-model constructed from ${\cal CP}^{3|4}$ supertwistor variables is proposed for the $AdS_5\times S^5$ superstring. At zero $AdS$ radius, free ${\cal N}$=4 d=4 super-Yang-Mills amplitudes are reproduced by topological amplitudes of the corresponding gauged linear sigma model where the closed superstring vertex operator for a trace of $k$ super-Yang-Mills fields is described by a boundary state with $k$ edges. After turning on a Fayet-Iliopoulis term in the sigma model with coefficient $R^2 =\sqrt{g^2_{YM} N}$, the topological amplitudes are claimed to reproduce the 't Hooft expansion of perturbative super-Yang-Mills amplitudes. Finally, this topological A-model is related to the usual $AdS_5\times S^5$ superstring in the pure spinor formalism.

\vskip .1in

\Date {June 2025}
\newsec{Introduction}

Proving the Maldacena conjecture that the superstring in an $AdS_5\times S^5$ background is dual to ${\cal N}$=4 d=4 super-Yang-MIlls has been a challenge because of the
complicated form of the $AdS_5\times S^5$ superstring worldsheet action. Using the pure spinor formalism, it was proposed that a topological A-model involving the supercoset
${{PSU(2,2|4)}\over {SU(2,2)\times SU(4)}}$ could describe the $AdS_5\times S^5$ superstring as a gauged linear sigma model \berklimit\berkvafa, but this model was not used to perform any calculations. 

In this paper, a topological A-model involving the supercoset ${{PSU(2,2|4)}\over {SU(1,2|4)}} = {\cal CP}^{3|4} $ will be proposed to describe the $AdS_5\times S^5$ superstring. This topological A-model is described by a gauged linear sigma model depending on an ${\cal N}$=4 d=4 supertwistor variable $\Lambda^\S$ and its complex conjugate $\Lb_\S$, together with its worldsheet superpartners and a U(1) N=(2,2) worldsheet gauge superfield. At zero radius, closed superstring vertex operators describing the trace of $k$ super-Yang-Mills fields will depend on the supertwistor variables in a manner similar to the Gaberdiel-Gopakumar vertex operators of \gopa, where $k$ is the ``vortex charge" which counts the contribution of the corresponding boundary state to the U(1) worldsheet instanton number. As in Ooguri-Vafa \vafa\ and Gaiotto-Rastelli \gaiotto, these boundary states ``fill the holes" of the thickened Feynman diagrams and provide the faces on the worldsheet in the 't Hooft expansion. At small radius, the topological amplitudes of these closed superstring vertex operators will be claimed to reproduce the corresponding  ${\cal N}$=4 d=4 super-Yang-Mills perturbative scattering amplitudes.

The topological A-model of this paper for the $AdS_5\times S^5$ superstring is very similar to previous twistor-string descriptions of  ${\cal N}$=4 d=4 super-Yang-Mills (and  ${\cal N}$=4 d=4 superconformal gravity) using B-model \wittenb, A-model \vafaa\ and open string \berktwistor\ theories, and to the recent proposal \gopa\ of a string dual to free ${\cal N}$=4 d=4 super-Yang-Mills. It would be very interesting to better understand the relation between these models, as well as the relation of the topological A-model of this paper to the description of \berkderive\ for the $AdS_5\times S^5$ superstring at small radius. Using the language of \gopatwo, the closed/open duality in this paper is an $F$-type duality where the closed superstring vertex operators get mapped to faces of the open string diagram. And the closed/open duality of \berkderive\ is a $V$-type duality where the closed superstring vertices are mapped to vertices of the open string diagram. So the two descriptions of the $AdS_5\times S^5$ superstring at small radius might be an example of open-closed-open triality as described in \gopatwo.

In section 2 of this paper, the topological A-model and corresponding gauged linear sigma model will be constructed and the boundary states for closed string vertex operators will be defined. In section 3, the topological amplitudes will be shown at zero radius to reproduce free super-Yang-Mills amplitudes and will be claimed to reproduce at small radius the 't Hooft expansion of perturbative super-Yang-Mills amplitudes. And in section 4, the topological A-model at finite radius will be related to the $AdS_5\times S^5$ superstring worldsheet action using the pure spinor formalism.

\newsec{Topological A-model}

The gauged linear sigma model for ${\cal CP}^{3|4} $ is constructed from a real N=(2,2) worldsheet superfield $V(\k^\pm, \kb^\pm)$ for the U(1) gauge multiplet and four bosonic and four fermionic worldsheet N=(2,2) chiral and antichiral superfields, $\Phi^\S(\k^+, \kb^+)$ and $\Phib_\S (\k^-, \kb^-)$, for the matter multiplets where $\S = (\a, \ad, j, \jb)$ is a $U(2,2|4)$ index, $(\a, \ad) = 1$ to 2 are $SU(1,1)$ indices, $(j, \jb)=1$ to 2 are $SU(2)$ indices, and $\k^\pm$ and $\kb^\pm$ are the left and right-moving Grassmann variables. The $\k^\pm=\kb^\pm=0$ components of $(\Phi^\S, \Phib_\S)$ will be called $(\L^\S, \Lb_\S)$.

The worldsheet action is 
\eqn\action{S = \int d^2 z \int d^2 \k^+ \int d^2 \k^- (\Phib_\S e^V \Phi^\S -  R^2 ~ V)}
where $R^4 =g^2_{YM} N$ is the 't Hooft coupling. The equation of motion $R^2 = \Phib_\S e^V \Phi^\S $ implies that when $R^2  \neq 0$, $V$ can be integrated out to give the
${\cal CP}^{3|4} $ worldsheet action 
\eqn\actiontwo{S = R^2 \int d^2 z \int d^2 \k^+ \int d^2 \k^- ~\log (\Phib_\S\Phi^\S)}
where $(\Phib_\S\Phi^\S)$ is non-vanishing so that the U(1) gauge symmetry is spontaneously broken.

However, when $R^2 =0$,  $\Phib_\S\Phi^\S=0$ and the U(1) gauge symmetry is unbroken.\foot{ In \gopavafa\vafa, the gauged linear sigma model at $R^2=0$ developed both a ``Higgs branch" where the gauge symmetry was spontaneously broken and a ``Coulomb branch" where the gauge symmetry was unbroken. In this paper, it will be assumed that only the Coulomb branch is present at $R^2 =0$.} After breaking the $U(2,2|4)$ to $U(1,1|2)\times U(1,1|2)$, $\Phib_\S \Phi^\S=0$ can be solved by defining $\Phi^\Ib = \Phib_I =0$ where
$\S=(I,\Ib)$ is a $U(2,2|4)$ index and $I=(\a, j)$ and $\Ib = (\ad, \jb)$ are
$U(1,1|2)$ indices. (As in harmonic superspace, complex conjugation will be defined such that $\Phib_\Ib$ is the complex conjugate of $\Phi^I$.) All other solutions of $\Phib_\S \Phi^ \S=0$
can be obtained by performing a $PSU(2,2|4)$ rotation on the solution $\Phi^\Ib = \Phib_I =0$. In the rest of this section, we will only discuss $R^2=0$ and will show how to describe closed string vertex operators for traces of free super-Yang-Mills fields using the topological A-model described by 
\eqn\actionfree{S_0 = \int d^2 z \int d^2 \k^+ \int d^2 \k^- ~\Phib_\S e^V \Phi^\S.} 

The choice $\Phi^\Ib = \Phib_I =0$ describes the half-BPS state annihilated by the $PSU(2,2|4)$ generators that leave $\Phi^\Ib = \Phib_I =0$ invariant, and
other half-BPS states can be described by modifying the solution of $\Phib_\S \Phi^\S=0$ to $\Phi^\Ib = i X^\Ib_I \Phi^I$ and $\Phib_I = -i \Phib_\Ib X^\Ib_I $ where $X^\Ib_I$ is a hermitian $(2|2)\times (2|2)$ super-matrix. Half-BPS states are also characterized by their conformal weight $k$, which at zero 't Hooft coupling corresponds to the number of super-Yang-Mills fields in the trace. Using the description of closed string vertex operators as boundary states as in \grone\gaiotto, the value of $k$ will be identified with the contribution of this boundary state to the instanton number of the U(1) worldsheet gauge field. This identification is natural since the number of $\Phi^I$ modes in the BRST cohomology of the topological A-model is related to the U(1) worldsheet instanton number on the surface \phases.

More explicitly, each closed superstring vertex operator dual to the trace of $k$ super-Yang-Mills fields will be described by a boundary state with $k$ edges of equal length and with the topology of a disk, i.e. a $k$-sided polygon. For an $n$-point scattering amplitude at genus $G$, the $n$ punctures of the genus $G$ Riemann surface will be expanded to $n$ boundary states as in \grone\gaiotto\ which are glued together so that the boundary states ``fill the holes" of the associated Feynman graphs in the 't Hooft expansion. So using the language of \gopatwo, this closed-open duality is F-type where the  closed string vertices are described by faces of the open string Feyman graphs. 

The ribbon graphs describing the edges between the boundary states will be assumed to be infinitesimally thin, and 
the values of $(\Phi^I, \Phib_\Ib)$ and the U(1) gauge field $V$ on the two sides of each edge will be assumed to be equal up to a U(1) gauge transformation, i.e. 
 \eqn\gauge{\Phi^{I(r)}(\s)= e^{i\xi_{rs}(\s)}\Phi^{I(s)}(\s), \quad 
\Phib_\Ib^{(r)}(\s)= e^{-i\xi_{rs}(\s)}\Phib_\Ib^{(s)}(\s),}
 $$A_z^{(r)}(\s) = A_z^{(s)}(\s)+ \partial_z\xi_{rs}(\s), \quad 
A_\zb^{(r)}(\s) = A_\zb^{(s)}(\s)+ \partial_\zb\xi_{rs}(\s),$$
where $(r)$ and $(s)$ label the fields on the adjoining faces $r$ and $s$, and $\s=0$ to $2\pi$ is the position on the edge between the two faces. 
Furthermore, it will be required that $\Phi$ and $\Phib$ are single-valued at all vertices (i.e. the endpoints of each edge where $\s=0$ or $\s=2\pi$), e.g.
\eqn\vertexcondition{\Phi^{I(r)} = \Phi^{I(s)} = \Phi^{I(t)} \quad {\rm and}\quad
\Phib_\Ib^{(r)} = \Phib_\Ib^{(s)} = \Phib_\Ib^{(t)}}
at the vertex where faces $r$, $s$ and $t$ meet. This implies that $\xi=0$ (mod 2$\pi$) at all vertices, e.g. $\xi_{rs} = \xi_{st} = \xi_{tr}=0$ (mod $2\pi$) at the point where faces $r$, $s$ and $t$ meet. 

Finally, it will be assumed that when moving clockwise around the $r^{\rm th}$ boundary state, $\int_0^{2\pi} d\s (A_\s^{(r)} - A_\s^{(s)}) = \int_0^{2\pi} d\s \p_\s\xi_{rs} =2\pi$ for each edge. So if the ``vortex charge" of a face is defined as the change in ${1\over{2\pi}}\xi$ when you go clockwise around the face and return to the same position, the ``vortex charge" of the $r^{\rm th}$ boundary state is equal to the number of its edges $k_r$.
Note that the total U(1) instanton number $E={1\over{2\pi}}\int d^2 z (\p_z A_\zb - \p_\zb A_z)$ of the surface is $E = \half \sum_r  k_r$ where the factor of $\half$ is because each edge appears on two faces. 

Each edge $E^{(r,a)}$ of the $r^{\rm th}$ face for $a=1$ to $k_r$ will describe a super-Yang-Mills state, and the state will be identified by the vertex operator 
\eqn\vertexop{V^{(r, a)}  = \exp [\pm  ~\Lb^{(r,a)}_\Ib (\s=\pi) ~(X^{(r,a)})^\Ib_I~ \L^{I(r,a)}(\s=\pi)]} 
where $(\L^{I(r,a)}(\s=\pi), \Lb^{(r,a)}_\Ib(\s=\pi))$ is the value of
$(\Phi^{I}, \Phib_\Ib)$ at $\k^\pm = \kb^\pm=0$ at the midpoint of the edge $E^{(r,a)}$.
Using the graded commutation relations of $\L^\S$ with $\Lb_\S$, one finds that when acting on $V^{(r,a)}$, 
$(\L^\Ib, \Lb_I)$ satisfy\foot{The choice of the $\pm$ sign in \vertexop\ depends if moving clockwise around the $r^{\rm th}$ face increases or decreases $\s$ on the edge. This is because the action of
\actionfree\ implies that at equal time $\tau$, one has the graded commutation relation
\eqn\graded{[\L^\S (\s, \tau), \Lb_\Psi (\s', \tau)] = i\d^\S_\Psi
\Theta (\s - \s')}
where $\Theta (\s-\s') = \pm 1$ depending if $\s-\s'$ is positive or negative. The equations $(\p_\tau +i \p_\s)\L^\S = (\p_\tau -i \p_\s)\Lb_\S =0$ can be used in deriving \graded\ since the topological A-model restricts to holomorphic solutions of $\Phi^\S$ and antiholomorphic solutions of $\Phib_\S$. So if one acts in a clockwise direction with $(\L^\Ib(\s), \Lb_I(\s))$ on the vertex operator of \vertexop, one will obtain the desired relation if one chooses the $+$ sign in \vertexop\ when moving in the clockwise direction increases the value of $\s$, and chooses the $-$ sign if moving in the clockwise direction decreases the value of $\s$.}
\eqn\sat{\L^\Ib = i (X^{(r,a)})_I^\Ib \L^I, \quad \Lb_I = -i \Lb_\Ib (X^{(r,a)})^\Ib_I .
 }
Since different choices of $X^{(r,a)}$ correspond to different soluctions of  $\Phib_\S \Phi^\S=0$, the vertex operator of \vertexop\ can be understood as choosing the $D$-brane boundary condition of edge $E^{(r,a)}$ for the dual open string theory. Note that the super-Yang-Mills state described by \vertexop\ and \sat\ can be expressed in harmonic superspace as the superfield
\eqn\harm{W(X^\Ib_I)= \phi_+(x^\ad_\a)+ y^\jb_j ~\phi^j_\jb(x^\ad_\a) + \e_{\jb\overline k}\e^{jk} y^\jb_j y^{\overline k}_k~ \phi_- (x^\ad_\a) + ...}
where $(\phi_+, \phi^j_\jb, \phi_-)$ are the 6 scalars of ${\cal N}=4$ d=4 super-Yang-Mills, $(x^\ad_\a, y^\jb_j)$ are the 8 bosonic components of $X^\Ib_I$, and ... depends on the 8 fermionic components of $X^\Ib_I$.

\newsec{ Topological amplitudes}

After performing the $A$-twist of the topological model, the functional integral over $\Phi^I$ and $\Phib_\Ib$ reduces to an ordinary integral over holomorphic and antiholomorphic solutions of 
$\overline \nabla\L^I =0$ and $\nabla\Lb_\Ib=0$. On a surface with worldsheet instanton number $E$, one can choose the U(1) field strength to be concentrated at $E$ distinct points on the worldsheet as $\p_z A_\zb - \p_\zb A_z = \sum_{k=1}^E \delta^2 (z-z_k)$. $\overline\nabla \L^I=0$ then implies that $\L^I$ can have simple poles at $z=z_k$, so  $\L^I$ is determined by the residues of these poles together with its value at any other fixed point on the surface. So on a surface of U(1) instanton number $E$, the functional integral reduces to an ordinary integral over $4(E+1)$ bosonic and $4(E+1)$ fermionic variables parameterizing the solutions of $\overline\nabla\L^I=\nabla\Lb_\Ib=0$ \phases. Since the surface is constructed by gluing together the boundary states described in the previous section with $k_r$ edges, the surface will contain $E=\half \sum_r k_r$ edges and it will be convenient to choose $4E$ bosonic and $4E$ fermionic variables to be the values of $(\L^I,  \Lb_\Ib)$ at the midpoints of each edge. The remaining 4 bosonic and 4 fermionic variables will be chosen to be the value of $(\L^I,  \Lb_\Ib)$ at any other point $(z_0, \zb_0)$ on the worldsheet.

The topological amplitude prescription at zero 't Hooft coupling for $n$ external closed string states dual to the trace of $k_r$ super-Yang-Mills fields for $E = \half \sum_{r=1}^n k_r$ is then given by 
\eqn\ampfree{{\cal A}_0 = \sum_{G=0}^\infty N^{2-2G} \sum_{S_G} \prod_{k=0}^E \int d\L^{I(k)} d\Lb_\Ib^{(k)} \prod_{r,a} V^{(r,a)} }
where $G$ is the genus of the worldsheet, $N$ is the number of colors of the super-Yang-Mills fields, $S_G$ labels the different ways one can glue together the $n$ faces to obtain a genus $G$ surface, $(\L^{I(k)}, \Lb_\Ib^{(k)})$ for $k=0$ to $E$ labels the values of $(\L^I, \Lb_\Ib)$ at the point $(z_0,\zb_0)$ and at the midpoints of the $E$ edges, and
$V^{(r,a)}$ are the $2E$ vertex operators of \vertexop\ located at the midpoints on both sides of each edge. Note that $N^{-1}$ plays the role of the genus coupling constant, and as expected for a topological string, the integral over genus $G$ worldsheet moduli has been replaced by a discrete sum $S_G$.

Using the form of $V^{(r,a)}$ in \vertexop, it is trivial to do the integral over $\prod_{k=1}^K \int d\L^{I(k)} d\Lb_I^{(k)}$ to obtain the amplitude 
\eqn\ampfreetwo{{\cal A}_0 = \sum_{G=0}^\infty N^{2-2G} \sum_{S_G} \int d\L^{I(0)} d\Lb_I^{(0)} \prod_{k=1}^E sdet^{-1} (X^{\Ib (k1)}_I - X^{\Ib (k2)}_I)}
where $X^{\Ib (k1)}_I$ and $X^{\Ib (k2)}_I$ come from the two vertex operators for the super-Yang-Mills states on the two sides of edge $k$. Note that these two vertex operators have opposite choice of $\pm$ sign since going clockwise around one face adjoining edge $k$ is going anticlockwise around the other face adjoining this edge. The final integral over
the four bosonic and fermionic zero modes of $\int d\L^{I(0)} d\Lb_I^{(0)}$ in \ampfreetwo\ gives 
$(\infty \times 0)^4$ since nothing depends on these zero modes. After regularizing this overall factor of $(\infty \times 0)^4$ to 1, one obtains 
\eqn\ampfreethree{{\cal A}_0 = \sum_{G=0}^\infty N^{2-2G} \sum_{S_G}  \prod_{k=1}^E sdet^{-1} (X^{\Ib (k1)}_I - X^{\Ib (k2)}_I)}
which agrees with the standard expression in harmonic superspace for the scattering amplitude at zero 't Hooft coupling of $n$ gauge-invariant super-Yang-Mills operators since, as shown in \howe, the propagator in harmonic superspace for super-Yang-Mills superfields, $W(X^{(1)})$ and $W(X^{(2)})$ of \harm, is $sdet^{-1} (X^{(1)} - X^{(2)})$.

Turning on the 't Hooft coupling corresponds to adding the term 
$$-R^2 \int d^2 z \int d^2 \k^+ \int d^2 \k^-    ~ V
 = -R^2 \int d^2 z ~D$$
 to the worldsheet action
of \actionfree\ where $D$ is the auxiliary field in the U(1) gauge multiplet. Since $D+ (\p_z A_\zb - \p_\zb A_z)$ is BRST-trivial \phases, adding the term $-R^2 \int d^2 z~ D$ shifts the action by  $2\pi R^2 E$ where
$E ={1\over{2\pi}} \int d^2 z (\p_z A_\zb - \p_\zb A_z)$ is the number of edges, and modifies the equation $\Phib_\S \Phi^\S =0$ to $\Phib_\S \Phi^\S =R^ 2$. It will be assumed that for small $R^2$, the modification to $\Phib_\S \Phi^\S =R^ 2$ can be reproduced by adding a vertex operator $R^2 \int d^2 z ~U$ to the action where  $U$ is a $PSU(2,2|4)$-invariant vertex operator in the BRST cohomology which deforms the topological A-model action of \actionfree.

The boundary state described by $U$ can in principle contain an arbitrary number of edges and will be denoted $U=U_0 + U_1 + U_2 + ...$ where $U_k$ contains $k$ edges. However, it will now be claimed that $PSU(2,2|4)$ invariance implies that only boundary states with 2 or 3 edges will contribute. Since each edge corresponds to a super-Yang-Mills field, this is consistent
with the fact that there are no local $PSU(2,2|4)$-invariant couplings of super-Yang-Mills fields with more than 3 fields. (Note that the usual quartic coupling
in the super-Yang-Mills action can be expressed as a cubic coupling by introducing auxiliary fields.)

The boundary state corresponding to $U_2$ with two edges is described by the vertex operator
\eqn\vutwo{U_2 = \exp ( +\Lb^{(1)}_\Ib ~X^\Ib_I~ \L^{I(1)}) \exp ( -\Lb^{(2)}_\Ib ~X^\Ib_I~ \L^{I(2)})}
where the opposite signs in the two exponents is because going clockwise around the two edges point in opposite directions. For this vertex operator to be
$PSU(2,2|4)$-invariant, one can either integrate it over all values of $X^\Ib_I$ or require that it is independent of $X^\Ib_I$. Integrating \vutwo\ over $X^\Ib_I$ produces an ill-defined expression, however, requiring that \vutwo\ is independent of $X^\Ib_I$ is easily seen to imply that $\Lb^{(1)}_\Ib = \Lb^{(2)}_\Ib$ and $\L^{I(1)} = \L^{I(2)}$ up to a U(1) gauge transformation. So gluing this boundary state to the edges of two other boundary states forces the values of $(\L^I, \Lb_\Ib)$ on the edges of these two other states
to coincide up to a U(1) gauge transformation. Therefore, gluing in the boundary state $R^2 \int d^2 z U_2$ simply multiplies each edge on the original surface by a factor of $c R^2$ where $c$ is some undetermined constant. So after gluing an arbitrary number of boundary states coming from pulling down factors of $R^2 \int d^2 z U_2$ in the action, the total effect on the scattering amplitude is to multiply all edges by a factor of $\exp (c R^2)$.

The boundary state corresponding to $U_3$ with three edges is described by the vertex operator
\eqn\vuthree{U_3 = \exp ( +\Lb^{(1)}_\Ib ~X^\Ib_I~ \L^{I(1)}) \exp ( +\Lb^{(2)}_\Ib ~X^\Ib_I~ \L^{I(2)})\exp ( -\Lb^{(3)}_\Ib ~X^\Ib_I~ \L^{I(3)})}
where we have assumed that two of the exponents have a $+$ sign and the third exponent has a $-$ sign. (The generalization of this analysis to other choices of the $\pm$ sign should be obvious.) In this case, $PSU(2,2|4)$ invariance implies up to U(1) gauge transformations that either
\eqn\either {\L^{I(1)} = \L^{I(2)} = \L^{I(3)}, \quad   \Lb^{(1)}_\Ib + \Lb^{(2)}_\Ib - \Lb^{(3)}_\Ib =0, \quad {\rm or}}
$$\Lb^{(1)}_\Ib = \Lb^{(2)}_\Ib = \Lb^{(3)}_\Ib, \quad   \L^{I(1)} + \L^{I(2)} - \L^{I(3)} =0.$$
After gluing the boundary state $U_3$ to the edges of three other boundary states, one finds that the first case in \either\ corresponds to the cubic super-Yang-Mills vertex with $+1$ ``bonus U(1) charge" whereas the second case corresponds to the cubic super-Yang-Mills vertex with $-1$ ``bonus U(1) charge" \wittenb\berktwistor. Note that ``bonus U(1) charge" is the U(1) of U(4) which is not an SU(4) R-symmetry, e.g. the self-dual and anti-self-dual gluons carry $+1$ and $-1$ charge, the chiral and antichiral gluinos carry $+\half$ and $-\half$ charge, and the six scalars carry zero charge.

For boundary states corresponding to $U_k$ with more than three edges, one can construct a similar vertex operator to \vuthree\ and one finds that $PSU(2,2|4)$ invariance implies that either
\eqn\eithertwo {\L^{I(1)} = \L^{I(2)} = ... = \L^{I(k)}, \quad   \Lb^{(1)}_\Ib \pm \Lb^{(2)}_\Ib \pm ... \pm \Lb^{(k)}_\Ib =0, \quad {\rm or}}
$$\Lb^{(1)}_\Ib = \Lb^{(2)}_\Ib = ... =  \Lb^{(k)}_\Ib, \quad   \L^{I(1)} \pm \L^{I(2)} \pm ... \pm \L^{I(k)} =0.$$
However, these boundary states contribute measure zero to the scattering amplitude since they can only contribute if the 4-momenta $(p_1^{\a \ad}, ..., p_k^{\a\ad})$ of the super-Yang-Mills states on the $k$ edges which are glued to this boundary state satisfy $p_r \cdot p_s =0$ for all $1\leq r,s \leq k$.

Finally, the boundary state $U_1$ with one edge does not contribute since independence of $X^\Ib_I$ implies either $\L^I =0$ or $\Lb_\Ib=0$.
And the boundary state $U_0$ with no edges does not contribute since it decouples from all other boundary states.

After including the contribution of $-R^2 \int d^2 z D =  2\pi R^2 E - R^2 \int d^2 z (U_2 + U_3)$ to the worldsheet action, the scattering amplitude ${\cal A}_0$ of 
\ampfreethree\ is modified to 
\eqn\ampint{{\cal A}_R = \sum_{G=0}^\infty N^{2-2G} ~ e^{- 2\pi R^2 E} e^{c R^2 E} (R^2)^{F'}~\sum_{S_G} \prod_{r'=1}^{F'} \int dX^{\Ib (r')}_I \prod_{k=1}^E sdet^{-1} (X^{\Ib (k1)}_I - X^{\Ib (k2)}_I) }
where  the factor of $e^{c R^2 E}$ comes from the contribution of $U_2$, $F'$ is the number of boundary states coming from $U_3$, and $X^{\Ib (r')}_I$ is the value of $X^\Ib_I$ appearing in the vertex operator of \vuthree\ for the $
{r'}^{\rm th}$ $U_3$ boundary state. The integral over $\prod_{r'=1}^{F'} \int dX^{\Ib (r')}_I$ will in general lead to divergences and need to be regularized as in the usual Feynman diagram computations.
To compare \ampint\ with the usual super-Yang-Mills perturbative amplitude, multiply each of the external vertex operators with $k$ edges by a factor of
$R^{2k-4}$. After rescaling the external operators in this manner, using that each face in $F'$ has 3 edges, and assuming that $c=2\pi$, the amplitude of \ampint\ can be expressed as
\eqn\ampinttwo{{\cal A}_R = \sum_{G=0}^\infty N^{2-2G}~  (R^4)^{E-F}~ \sum_{S_G}  \prod_{r'=1}^{F'} \int dX^{\Ib (r')}_I  \prod_{k=1}^E sdet^{-1} (X^{\Ib (k1)}_I - X^{\Ib (k2)}_I)}
$$= \sum_{G=0}^\infty (N R^{-4})^{2-2G} ~  (R^4)^V~\sum_{S_G}  \prod_{r'=1}^{F'} \int dX^{\Ib (r')}_I   \prod_{k=1}^E sdet^{-1} (X^{\Ib (k1)}_I - X^{\Ib (k2)}_I)$$
$$= \sum_{G=0}^\infty (g^2_{YM})^{2G-2}~ (g^2_{YM} N)^V~ \sum_{S_G}  \prod_{r'=1}^{F'} \int dX^{\Ib (r')}_I   \prod_{k=1}^E sdet^{-1} (X^{\Ib (k1)}_I - X^{\Ib (k2)}_I) ,$$
where $F$ is the the total number of faces and $V$ is the total number of vertices. Since the vertices in this ``$F$-dual" description correspond to faces of the usual super-Yang-Mills Feynman diagrams, ${\cal A}_R$ agrees with the standard 't Hooft expansion of super-Yang-Mills perturbative amplitudes.

\newsec {Relation to $AdS_5 \times S^5$ superstring}

As in any Type IIB supergravity background, the worldsheet action using the pure spinor formalism in an $AdS_5 \times S^5 $ background is constructed from the 10 $x^m$ variables and 32 $(\t^\a, \tb^\a)$ variables of N=2 d=10 superspace together with left and right-moving pure spinor ghost variables $(\l^\a, \lb^\a)$
carrying 22 independent bosonic degrees of freedom. As was shown in \berklimit\berkvafa, it is natural in an $AdS_5\times S^5$ background to combine the 22 pure spinor ghost variables $(\l^\a, \lb^\a)$ and the 10 $x^m$ spacetime variables into the
32 variables $(Z^A_J, \Zb_A^J)$ transforming as $SO(4,2)\times SO(6)$ unconstrained spinors where $A=1$ to 4 is an $SU(2,2)$ index and $J=1$ to 4 is an $SU(4)$ index. Under an $A$-twisted N=(2,2) worldsheet supersymmetry, these bosonic twistor-like variables $(Z^A_J, \Zb_A^J)$ are the worldsheet superpartners of $(\t^A_J, \tb_A^J)$ and combine into the chiral and anti-chiral N=(2,2) superfields $\T^A_J(\k^+, \kb^+)$ and $\Tb_A^J(\k^-, \kb^-)$. The resulting topological A-model action is \berklimit
\eqn\sfour{S = R^2 \int d^2 z d^2 \k^+ d^2 \k^- ~Tr [log (1 + \Tb\T)],}
which can also be expressed as a gauged linear sigma model action based on the supercoset ${{PSU(2,2|4)}\over{SU(2,2)\times SU(4)}}$ \berkvafa.

In \berkpert, it was shown that this topological A-model can be obtained by gauge-fixing a ${\cal G}/{\cal G}$ principal chiral sigma model where 
${\cal G} = PSU(2,2|4)$. The action for the ${\cal G}/{\cal G}$ model is
\eqn\actg{ S = R^2 \int d^2 z Tr [(g^{-1} \p g - A) (g^{-1} \pb g - \Ab)]}
where $g$ takes values in $PSU(2,2|4)$ and $(A,\Ab)$ is a worldsheet gauge field taking values in the $PSU(2,2|4)$ Lie algebra. To obtain \sfour\ from \actg,
use the local $PSU(2,2|4)$ symmetry to gauge-fix the 30 bosonic components of $g$ and to gauge-fix to zero the 32 fermionic components $A^A_J$ and $\Ab_A^J$.
Gauge-fixing $g$ does not introduce propagating ghosts, but gauge-fixing $A^A_J= \Ab_A^J=0$ produces the Fadeev-Popov ghosts $(Z^A_J, \Zb_A^J)$ and
their conjugate momenta which are the worldsheet superpartners of $(\t^A_J, \tb_A^J)$ in \sfour.

Moreover, it was shown in \berksimp\ that the original $AdS_5\times S^5$ worldsheet action of \berkads\ using the pure spinor formalism can also be obtained by gauge-fixing
the  ${\cal G}/{\cal G}$ principal chiral sigma model action of \actg. To obtain the pure spinor worldsheet action based on the Metsaev-Tseytlin supercoset
${PSU(2,2|4)}\over{SO(4,1)\times SO(5)}$, one uses the $SO(4,1)\times SO(5)$ subset of local $PSU(2,2|4)$ symmetries to gauge fix the $SO(4,1)\times SO(5)$ subset of $g$. The remaining 10 bosonic and 32 fermionic local $PSU(2,2|4)$ symmetries are used to gauge-fix 5 bosonic and 16 fermionic
components of $A$ and 5 bosonic and 16 fermionic components of $\Ab$. Using the ``extended pure spinor formalism" of Aisaka and Kazama \aisaka, this gauge-fixing
of $(A,\Ab)$ was argued to produce the correct set of constrained pure spinor ghosts. 

Finally, the ${\cal CP}^{3|4}$ sigma model action of \actiontwo\ in this paper can also be obtained by gauge-fixing the ${\cal G}/{\cal G}$ principal chiral sigma model action of \actg. In this case, one uses an $SU(1,2|4)$ subgroup of the local $PSU(2,2|4)$ symmetries to gauge away 24 bosonic and 24 fermionic
component of $g$. The remaining 6 bosonic and 8 fermionic local symmetries are used to gauge-fix $A^1_{\S'} = \Ab_1^{\S'} =0$ where
$\S'$ ranges over all $PSU(2,2|4)$ indices except for the first one. In this gauge, the remaining 6 bosonic and 8 fermionic components of $g$ describe the
${\cal CP}^{3|4}$ worldsheet variables and the Fadeev-Popov ghosts coming from gauge-fixing $(A, \Ab)$ produce the worldsheet N=(2,2) superpartners of
these ${\cal CP}^{3|4}$ variables.

So it has been shown that the topological A-model of this paper and the $AdS_5\times S^5$ superstring action of \berkads\ are related by different gauge-fixings of the 
${\cal G}/{\cal G}$ principal chiral sigma model of \actg. Since any superstring background can in principle be obtained by deforming the $AdS_5\times S^5$ at small radius, the ${\cal G}/{\cal G}$ principal chiral sigma model may be a more natural starting point for superstring theory than the usual superstring action in a flat d=10 background. A similar proposal to use a topological sigma model as an ``unbroken phase" of string theory was discussed in \wittop.

\vskip 10pt
{\bf Acknowledgements:}
I would like to thank Matthias Gaberdiel, Rajesh Gopakumar, Juan Maldacena, Cumrun Vafa and Pedro Vieira for useful discussions, and 
CNPq grant 311434/2020-7
and FAPESP grants 2016/01343-7, 2021/14335-0, 2019/21281-4 and 2019/24277-8 for partial financial support.

\footatend\vfill\supereject\immediate\closeout\rfile\writestoppt
\baselineskip=14pt\centerline{{\bf References}}\bigskip{\frenchspacing%
\parindent=20pt\escapechar=` \input refs.tmp\vfill\eject}\nonfrenchspacing

\end
\bye